\documentclass[preprint,aps,pra]{revtex4}
\newcommand{\eqref}[1]{(\ref{#1})}
\def\sn{{\rm sn\,}}
\def\cn{{\rm cn\,}}
\def\dn{{\rm dn\,}}

\begin{document}

\title{\bf{ Calculation of Band Edge Eigenfunctions and
Eigenvalues of Periodic Potentials through the Quantum Hamilton -
Jacobi Formalism \footnote{ This paper is dedicated to the memory of
  Prof. R. A. Leacock.}}}
\author{
  S. Sree Ranjani$^{a}$ \footnote{akksprs@uohyd.ernet.in},
  A.K. Kapoor$^{a}$ \footnote{akksp@uohyd.ernet.in} and
  P.K. Panigrahi$^{b}$ \footnote{prasanta@prl.ernet.in}}
 \affiliation{$^a$ School of Physics, University of Hyderabad, Hyderabad 500 046, India\\
$^b$ Physical Research Laboratory Navrangpura, Ahmedabad, 380009,
India}

%\maketitle

\begin{abstract}
  We obtain the band edge eigenfunctions and the
eigenvalues of solvable periodic potentials using the quantum
Hamilton - Jacobi formalism. The potentials studied here are the
Lam{\'e} and the associated Lam{\'e} which belong to
the class of elliptic potentials. The formalism requires an assumption
about the singularity structure of the quantum momentum
function $p$, which satisfies the Riccati type quantum Hamilton -
Jacobi  equation, $ p^{2} -i \hbar \frac{d}{dx}p = 2m(E- V(x))$ in
the complex $x$ plane. Essential use is made of suitable conformal
transformations, which leads to the eigenvalues and the eigenfunctions
corresponding to the band edges in a simple and straightforward
manner. Our study reveals interesting features about the
singularity structure of $p$, responsible in yielding the band
edge eigenfunctions and eigenvalues.
\end{abstract}

\maketitle

\section{Introduction}

 As is well - known, periodic potentials play a significant role in
condensed matter physics. Recently, optical lattices have
manifested in Bose -  Einstein condensates (BEC),$^{\ref{opl}}$
leading to the analysis of the excitation spectra in such systems.
The energy spectrum of periodic potentials is unique due to the
existence of energy bands and the solutions of the
Schr$\ddot{o}$dinger equation have the Bloch form:
\begin{equation}
\psi(x) = u(x) \exp(ik.x).           \label{e1}
\end{equation}
Here $u(x)$ has the periodicity of the potential function. These
features of the periodic potentials are usually illustrated using
the Kronig - Penney model,$^{\ref{kit}}$ in which the crystal
lattice is approximated by a periodic array of square wells.
Imposing the Bloch condition on the solutions of the Kronig-Penney
model, one is led to a transcendental equation, which has to be
solved in order to obtain the band edge solutions.

Of late, there has been considerable interest in the study$^{\ref{kuz} - \ref{osc}}$ and construction$^{\ref{khare},\ref{d2}}$ of
new periodic potentials. The periodic lattices in BEC has also
renewed interest in this area.$^{\ref{bec1}}$ The number of exactly
solvable (ES) potentials is very limited and in that, exactly
solvable periodic potentials constitute a very small number. Quite
sometime back it was shown by Scarf that, $ V(x) =  A/ \sin^{2}x$,
usually treated in a finite domain, can lead to band structure for
$ -1/4 < A < 0 $ $^{\ref{sc}}$. In this paper, we study the
Lam{\'e} and the associated Lam{\'e} potentials which belong to the 
class of elliptic potentials which are usually expressed in terms of the Jacobi
elliptic functions $ \sn (x,m),\cn (x,m)$ and $\dn (x,m)$. The parameter $m$
is known as the elliptic modulus whose value lies between 0 and 1. By
applying supersymmetric quantum mechanics $^{\ref{khare},\ref{d2}}$ and group theoretical techniques to the family of
elliptic potentials, many new ES and QES potentials have been constructed 
$^{\ref{asis},\ref{tka}}$. 

   The Lam{\'e} potential
\begin{equation}
V(x) = j(j+1)m \, \sn ^{2}(x,m),        \label{e2}
\end{equation}
with $j$ being an integer, has $2j+1$ bands followed by the continuum.
This potential has been  extensively studied$^{\ref{ars} - \ref{whit}}$. It is interesting to
note that the Schr$\ddot{o}$dinger equation with the Lam{\'e} potential
\begin{equation}
\frac{d^{2}\psi(x)}{dx^{2}} +\frac{\hbar^{2}}{2m}( E - j(j+1)m\, \sn ^{2}(x,m))\psi(x) =
0,   \label{e4}
\end{equation}
is a result of separating the  Laplace equation in the ellipsoidal
coordinates  and is
referred to as the Lam{\'e} equation$^{\ref{ars},\ref{whit}}$. This has unique
analytical properties$^{\ref{ars}}$, which render it
useful in the study of topics ranging from astrophysics $^{\ref{fin}
- \ref{bac}}$ to condensed matter physics$^{\ref{gli}}$. For example
one comes across Lam{\'e} equation in problems related to the
early  epoch of the universe like the distance red-shift relations
$^{\ref{kan}}$ and quantum vacuum fluctuations.$^{\ref{fin}}$
The Lam{\'e} equation occurs in the study of bifurcations
in chaotic Hamiltonian systems$^{\ref{brac1},\ref{brac2}}$. It was
shown by Finkel {\it et.al} $^{\ref{fin}}$ that the matter modes
equation, corresponding to the  most general inflation potential in
the Minkowski space reduces to the Lam{\'e} form.

In comparison, the associated Lam{\'e} potential
\begin{equation}
V(x) = pm \, \sn^{2}(x,m)+qm \frac{\cn^{2}(x,m)}{\dn^{2}(x,m)},   \label{e3}
\end{equation}
with $p =a(a+1)$ and $q = b(b+1)$, did not have a systematic study 
till now and has a few scattered
references in the mathematical literature. This potential is ES
when  $a = b =j$, with $j$ being an integer and QES when $ a \neq b$ 
with $a$, $b$ being real. Recently, a systematic
study of the band structure of this potential, for various cases
of $a$ and $b$, has been done by Khare and Sukhatme$^{\ref{uday}}$ and 
the algebraic aspects of the QES associated Lam{\'e}
potential have been investigated in$^{\ref{tur1} - \ref{alh}}$.

   The band edge eigenfunctions and the eigenvalues of both these potentials
have been found using the conventional techniques of solving the ordinary
differential equations$^{\ref{ars}, \ref{whit}}$. Here, we
propose an alternate method through the quantum Hamilton -
Jacobi (QHJ) formalism, where one can obtain the solutions,
without solving the differential equations explicitly. The QHJ
formalism was shown to yield bound state energies and wave functions
for a large number of ES models. The tools used  for our earlier
studies are sufficient to derive the results for the periodic
potentials also. This method requires making  a guess about the
singularity structure of the logarithmic derivative of the wave
function $\psi$ and using elementary tools from
complex variables to fix the QMF completely. Once the form of the QMF
is obtained getting the band edge wave functions and eigenvalues is
straightforward. 

In the next section, we give a brief summary of the QHJ formalism
and the steps involved in obtaining the solutions for any general
potential. In sec $III$, we show how one can obtain the form of
the wave functions for the Lam{\'e} potential for a general
$j$, followed by sec $IV$, where we take up a special case of $
j=2$ and obtain explicit band edge wave functions and energies. In
sec $V$, the form of the wave functions for the  general
associated Lam{\'e} potential (ES case) is  derived. Associated
Lam{\'e} potential, with $a = b = 1$, is worked out in
sec $VI$, followed by the last  section containing the concluding
remarks.

\section{Quantum Hamilton- Jacobi Formalism }

    The QHJ formalism, was initiated by Leacock and
Padgett$^{\ref{lea},\ref{pad}}$ in 1983 - 84. Using this formalism on a
host of ES models, Bhalla {\it et.al},$^{\ref{bh1} - \ref{bh4}}$
were successful in obtaining the corresponding energy eigenvalues.
The method when applied to the QES models, yielded the quasi
exactly solvability  condition$^{\ref{geo}}$. With a slight
modification of the method, it was shown that, one could obtain
the eigenfunctions and the eigenvalues of ES models simultaneously$^{\ref{sree}}$. The main aim of this present study is to demonstrate
the applicability of the QHJ formalism to the periodic potentials.
Since the QHJ formalism and its working are well studied in the
literature, we do not go into details here but give only the
necessary information. For details, the interested reader is
referred to the earlier works$^{\ref{lea} - \ref{sree}}$ and the
references there in.

    The main object of interest in the QHJ formalism is the
quantum momentum function (QMF) $p$, which is the logarithmic derivative of
the wavefunction $\psi(x)$,$^{\ref{lea}, \ref{pad}}$ apart from a factor of $-i\hbar$
\begin{equation}
p = -i \hbar \frac{d}{dx}(\ln\psi(x)).     \label{e5a} 
\end{equation} 
Hence, by obtaining the expression for the QMF one can obtain the expression
for the wavefunction. The QMF satisfies the Riccati equation which is referred
to as the QHJ equation,
\begin{equation}
p^{2}-i\hbar p^{\prime} = 2m(E-V(x)),       \label{e5}
\end{equation}
which is the special case of the general Riccati equation
\begin{equation}
A(x)p^{2} +B(x)p +C(x) +i \frac{dp}{dx}  = 0.    \label{e5b}
\end{equation}
Equation \eqref{e5}{} has the most convenient form to work with, in the QHJ
formalism. Hence for all the
cases studied through this method one tries to the bring the equation for
the QMF to the form of \eqref{e5}{}. In general, the solutions of Riccati
equation \eqref{e5b}{} has two types of
singularities, the fixed singularities and the moving  singularities. The fixed
singularities are determined by the singular points of $A(x)$, $B(x)$ and
$C(x)$. These appear in every solution and are independent of the initial 
conditions. Thus for the QHJ equation \eqref{e5}, the fixed singular points
originate from the potential. The other singular points {\it i.e}, the moving
singular points 
depend on the initial conditions and need not appear in every solution. It is
known that for the Riccati equation only poles can appear as moving
singularities $^{\ref{ince}}$.

   It is  known that, the wavefunction for the $n^{th}$ excited state has $n$
zeros. It is then seen from \eqref{e5}{} that correspondingly $p$ has $n$
moving poles in between the two  classical turning
points and the residue at each of this pole is $-i \hbar$. In general there
could be other moving poles of $p$ in the finite complex plane but one has
little information regarding their location.

   In our earlier studies of the ES and QES solvable models it was assumed
that the QMF has finite number of moving poles in the finite complex
plane,$^{\ref{geo},\ref{sree}}$ which turned out to be true for all the models
studied.  Therefore, for the study of the exactly solvable periodic
potentials, without losing generality, one can assume that, the QMF has finite
number of moving poles in the complex plane. It follows from our assumption
that the point at infinity is an isolated singularity. 

With this assumption on the singularity structure of the
QMF, we proceed to write the QHJ equation  \eqref{e5}{} as
\begin{equation}
q^2 + q^{\prime} = V(x) - E,      \label{e6}
\end{equation}
with $ \hbar = 2m = 1$, where $ q = \frac{d}{dx}\psi(x)$. The wavefunction in terms of $q$ will be
\begin{equation}
\psi(x) = \exp(\int{q(x)dx}).      \label{e7}
\end{equation}
 One can see from \eqref{e6}{} that the residue at the moving poles is unity.
In general it has been found useful to change  variable $x$ to
variable $t = f(x)$  so as to make the
coefficients of the Riccati equation rational. After a change of variable,
\eqref{e6} becomes
\begin{equation}
q^2 + F(t)\frac{dq}{dt} + E - \tilde{V}(t)  = 0,    \label{e7a}
\end{equation}
where $ F(t) = \frac{df}{dx}$ expressed as a function of $t$ and
$\tilde{V(t)}$ is the potential in terms of $t$. One can see that
the above equation does not have the convenient form of
\eqref{e6}{}. Hence to get \eqref{e7a} into the form of
\eqref{e6}, we perform a transformation from $q(x)$ to $\chi(t)$
as follows,
\begin{equation}
q = F(t)\phi  ,\,\,  \phi = \chi - \frac{1}{2}\frac{d}{dt}(\ln
F(t)).   \label{e9}
\end{equation}
Using the above transformations \eqref{e7a}{} becomes
\begin{equation}
\chi^{2} + \frac{d\chi}{dt} + \frac{E-\tilde{V(t)}}{F^{2}(t)} -
  \frac{1}{2} \left(\frac{F^{\prime\prime}(t)}{F(t)}\right) +
  \frac{1}{4} \left(\frac{F^{\prime}(t)}{F(t)} \right)^{2}  = 0,    \label{e8}
\end{equation}
which  is in the form of the Riccati equation. The residue at the
moving poles is unity. The fixed poles correspond to the zeros of
$ F(t)$. One can make use of \eqref{e8}{} instead of the
original QHJ equation for any general potential. In the next
section we show, by taking the explicit example of the general
Lam{\'e} potential, how one can obtain the form of the band edge wave
functions.

\section{General Lam{\'e} Potential}

  The QHJ equation in terms of $q$, for the Lam{\'e} potential is
\begin{equation}
q^2 + q^{\prime} + E -j(j+1)m\, \sn ^{2}(x,m) = 0.     \label{e10}
\end{equation}
To proceed further we need to bring the potential to a meromorphic
form, for which we do the following the change of variable
\begin{equation}
t = \sn(x,m),      \label{e11}
\end{equation}
with $\sn x = \sn (x+4 K(m))$. We would like to point out here that,
the Lam{\'e} equation 
can be written in five forms namely two algebraic, one
trigonometric, one Weierstrassian and one Jacobi$^{\ref{ars}}$, 
depending on the change of variable. Of all these change of 
variables we found \eqref{e11}, which gives the Jacobi form of this 
Lam{\'e} equation, to be best suited for our method. It enables 
one to write the QHJ equation in a form which can be 
easily analyzed using the QHJ formalism. Another added advantage of 
this particular choice is that, it maps half of the period
parallelogram $ 0 \leq x <2 K(m) $ 
of the Jacobi elliptic $\sn(x,m)$ function to the  
upper half complex plane.$^{\ref{chu}}$ One gets the equation for $\chi$ as
\begin{equation}
\chi^2 +\frac{d\chi}{dt} +\frac{(mt)^{2}+2m}{4(1-mt^{2})^2} +
\frac{t^{2}+2}{4(1-t^2)^2}
+\frac{2E-2j(j+1)mt^{2}-mt^{2}}{2(1-t^2)(1-mt^2)} = 0,  \label{e14}
\end{equation}
where
\begin{equation}
q = \sqrt{(1-t^2)(1-mt^2)}\phi ,\,\,  \phi = \chi
+\frac{1}{2}\left(\frac{mt}{1-mt^2} + \frac{t}{1-t^2} \right).   \label{e15}
\end{equation}
The following properties of the Jacobi elliptic functions$^{\ref{han}}$
were used.
\begin{equation}
\frac{d}{dx}\sn(x,m) = cn(x,m)dn(x,m)    \label{e12}
\end{equation}
and
\begin{equation}
\sn^{2}(x,m)+cn^{2}(x,m)=1,\,\,  \sn^{2}(x,m) + mdn^{2}(x,m) = 1.
\label{e13}
\end{equation}

For all our further calculations we will regard $\chi$ as the QMF
instead of $p$. Using \eqref{e14}, in place of the original QHJ
equation \eqref{e10}{}, one can obtain the expressions for the
wave functions by analyzing the singularity structure of $\chi$

{\bf Singularity  structure of $\chi$ :} Equation \eqref{e14}{} shows that $\chi$ has fixed poles at $t = \pm 1$
and $t=\pm 1/ \sqrt{m}$ and there are finite number of
moving poles in the complex plane. Hence we make an assumption
that the point at $\infty$ is an isolated singular point and that
there are no singularities of the QMF except for those
mentioned above. Therefore one can
write $\chi$, separating it into its singular and analytical
parts in the following form,
\begin{equation}
\chi =
\frac{b_{1}}{t-1}+\frac{b^{\prime}_{1}}{t+1}+\frac{d_{1}}{t-1/\sqrt{m}}+\frac{d^{\prime}_{1}}{t+1/\sqrt{m}}
+\left( \sum_{k=0}^{n}\frac{1}{t-t_{k}}\right) + Q(t). \label{e16}
\end{equation}
Here, $Q(t)$ is the analytic part of $\chi$ and the rational terms
represent the singular parts. Here, $b^{\prime}_{1}$,$b_1$,
$d^{\prime}_{1}$ and $d_{1}$ are the residues at $ t=\pm 1$ and $t
= \pm 1/\sqrt{m}$ respectively. From \eqref{e14}, one can see that
$\chi$ is bounded at infinity, which makes $Q(t)$ in \eqref{e16},
analytic and bounded at infinity. Hence from Louville's
theorem $Q(t)$ will be a constant $C$. The summation term in
\eqref{e16}{}, represents the singular part coming due to the
finite number of moving poles for which the residues are easily
found to be one and this term can be written as $
\frac{P^{\prime}_{n}}{P_{n}}$, 
where $P_{n} \equiv \displaystyle {\prod_{k=0}^{n}} (t - t_{k})$
is an $n^{th}$ degree polynomial. Thus \eqref{e16}{} becomes
\begin{equation}
\chi =
\frac{b_{1}}{t-1}+\frac{b^{\prime}_{1}}{t+1}+\frac{d_{1}}{t-1/\sqrt{m}}+\frac{d^{\prime}_{1}}{t+1/\sqrt{m}}
+\frac{P^{\prime}_{n}}{P_n} + C.    \label{e18}
\end{equation}
The residues $b_1$, $b_{1}^{\prime}$, $d_{1}$ and $d_{1}^{\prime}$
at the fixed poles can be determined by substituting the
Laurent expansion of $\chi $ around these poles. For example, to
calculate the residue at $t=1$, one expands $\chi$
\begin{equation}
\chi = \frac{b_1}{t -1} + a_{0} +a_{1}(t-1) +........ \label{e18a}
\end{equation}
Substituting the above equation in \eqref{e14}{} and  comparing
the coefficients of different powers of $(t-1)$, one gets a
quadratic equation in $b_1$ whose roots are
\begin{equation}
b_{1}= 3/4  ,\,\, 1/4.    \label{e18b}
\end{equation}
Similarly the
two values of the residues  at all the other fixed poles $
b^{\prime}_{1},d_1$ and $ d^{\prime}_{1}$ turn out to be
\begin{equation}
 3/4 ,\,\, 1/4.        \label{e18c}
\end{equation}
Note that, all the values of the residues are independent of the
potential parameter $j$. Hence, irrespective of the value of $j$
in the potential, the residues at the fixed poles will take only
the above values for any Lam{\'e} potential.

   It will be shown later that, the knowledge of the residues is
sufficient to obtain the wave function. Hence, at this stage, we
should check, which values of the residues, out of the two values,
give rise to acceptable`
wave functions. In the present case, it
turns out that both the values, i.e, $3/4$ and $1/4$ for each of
the residues, leads to acceptable wave functions. The only
requirement that restricts possible combinations of values is that
of parity. The fact that $\psi$ has definite parity implies that
$\chi(-t) = -\chi(t)$, which rules out the possibilities  $ b_1 \neq
b^{\prime}_1$ and $ d_1 \neq d^{\prime}_1$ and hence of all the possible
combinations of the values of $b_1$, $b_{1}^{\prime}$, $d_{1}$ and
$d_{1}^{\prime}$, the only ones which are accepted are those
with $ b_1 = b^{\prime}_1$ and $ d_1 = d^{\prime}_1$.

   In contrast to the above results, in our earlier study of ES boundstate
problems$^{\ref{sree}}$ and QES problems$^{\ref{geo}}$ one was lead
to a unique choice of the residues at the fixed poles when one
demanded that the wave function be square integrable. The other
values of the residue did not correspond to any physical solution and
were ruled out. In the case of potentials, which exhibit two phases of
SUSY, for different ranges of the potential parameters$^{\ref{sree}}$
both values of the residues were useful. One set of residues gave
physically acceptable solutions for exact SUSY case and the other
set gave for the broken SUSY. In the case of periodic potentials there is
no such restriction as square integrability and so there is no way of
ruling out one of the values. The reason that only one value of residue 
is acceptable for the bound state problems in one dimension can be
attributed to the fact that the bound state solutions are
non-degenerate. Whereas the band edge wavefunctions of the periodic
potentials are doubly degenerate,  we keep both the values of the residues.

{\bf Behaviour of $\chi$ at infinity :}   We have assumed that the point at infinity is an isolated
singularity. The form of \eqref{e14}{} suggests that $\chi$ is bounded
at $\infty$. Therefore, $\chi$ can be expressed as
\begin{equation}
\chi(t) = \lambda_0 +\frac{\lambda_1}{t} +\frac{\lambda_2}{t^2} +....
\label{e19}
\end{equation}
and the coefficients $\lambda_k$'s are fixed using the Riccati equation
\eqref{e14}{} and one gets
\begin{equation}
\lambda_0 = 0 ,    \label{e20}
\end{equation}
and two values for $\lambda_1$ as
\begin{equation}
\lambda_1 = j+1 ,\,\,  -j.    \label{e21}
\end{equation}
From \eqref{e21}{}, one obtains the leading behaviour of $\chi(t)$
at $\infty$ which is dependent on $j$. Equation \eqref{e18}{} gives the form
of $\chi$ which holds for all finite values of $t$. For large $t$,
comparing \eqref{e18}{} and \eqref{e19}{}, one gets
\begin{equation}
2b_{1} + 2d_{1}+n = \lambda_1.      \label{e22}
\end{equation}
As the left hand side is positive, it is clear that, only the
choice $\lambda_1 = j+1 $ is consistent with the above equation,
while the other value $ \lambda_1 = -j$ is ruled out. Thus from
\eqref{e22}{}, one gets, the expression for $n$ as
\begin{equation}
n = j+1 -2b_1 -2d_1 ,   \label{e23}
\end{equation}
which gives the degree of the polynomial $P_n$.

{\bf Explicit forms of the wavefunctions :} One can obtain the form of
the wave functions using \eqref{e7}{}. 
Substituting the relations given in \eqref{e15}{} in \eqref{e7}{},
one gets the expression for the wave function in terms of $\chi$
as,
\begin{equation}
\Psi(t) = \exp \int \left(\chi(t) +
\frac{1}{2}\left(\frac{mt}{1-mt^2}+\frac{t}{1-t^2}\right)\right)dt
\label{e24}
\end{equation}
\begin{equation}
             =  \exp \int
        \left(\frac{(1-4b_1)t}{2(1-t^2)}+\frac{(1-4d_1)mt}{2(1-mt^2)}
        +\frac{P^{\prime}_n}{P_n}\right)dt.    \label{e25}
\end{equation}
which when simplified and written in terms of the original variable
$x$, gives
\begin{equation}
\psi(x) = (\cn x)^{\alpha} (\dn x)^{\beta} P_{n}(\sn x),    \label{e26}
\end{equation}
where, $\alpha = \frac{4b_1 - 1}{2}$ and $\beta = \frac{4d_1
  -1}{2}$. For the sake of simplicity, the elliptic modulus $m$ of
the Jacobi elliptic functions will be suppressed here onwards. The
four different combinations of the residues (given as sets 1 to 4
) give rise to four different forms of the band edge wavefunctions as
listed in the table I.
%table I
%\begin{tabular}{c@{}|c@{}|c@{}|c|c@{}|c@{}|c@{}|p{0.25in}|}
%\hspace{1.5in}
%\begin{table}[p]
%\caption{Form and number of solutions for the general Lam{\'e} potential }
%\begin{tabular}{|c|c|c|c|c|c|c|}
%\hline
%\multicolumn{7}{|c|}
%   \\  \hline
%\multicolumn{1}{|c|}{$set$}
%&\multicolumn{1}{|c|}{$b_1$}
%&\multicolumn{1}{|c|}{$d_1$}
%&\multicolumn{1}{|c|}{n}
%&\multicolumn{1}{|c|}{$\psi$ in terms of x}
%&\multicolumn{2}{|c|}{Number of solutions} \\
%\cline{6-7}
%\multicolumn{1}{c|}{$\psi$ in terms of x}
%&&&&&\multicolumn{1}{|c|}{$j=2N+1$}
%&\multicolumn{1}{|c|}{$j=2N$ }
% \\ \hline
%$1$ & $1/4$ & $1/4$ & $j$ & $P_{j}(\sn x)$ & $N+1$ & $N+1$    \\
%$2$ & $3/4$ & $1/4$ & $j-1$ & $cnxP_{j-1}(\sn x)$ & $N+1$ & $N$  \\
%$3$ & $1/4$ & $3/4$ & $j-1$ & $dnxP_{j-1}(\sn x)$ & $N+1$ & $N$  \\
%$4$ & $3/4$ & $3/4$ & $j-2$ & $cnx dnxP_{j-2}(\sn x)$ & $N$ & $N$  \\
%\hline
%\end{tabular}
%\end{table}
%\clearpage
%\vskip0.35cm

    The parity constraint $\chi(-t) = -\chi(t)$, restricts the polynomial
$P_n (t)$ to have either only odd or only even powers of $t \equiv \sn
    x$. In the sixth and the seventh columns of table I, the number of linearly
independent solutions, for the two cases, $j$ being odd and $j$ being even are
given in terms of a positive integer $N$. For odd $j$, $N= (j-1)/2$
and for even $j$, $N = j/2$. In both the cases the total number of
solutions for a particular $j$ is equal to $2j +1$. It is easy to see that,
the four forms and the number of solutions, of a particular form
obtained here are in agreement with those already known$^{\ref{ars}}$. For a given set of $b_1 $and $d_1$, $n$ is fixed
using \eqref{e23}{} and the differential equation for the unknown
polynomial $p_n (t)$ can be obtained by substituting $\chi(t)$
from \eqref{e18}{} in \eqref{e14}{}, which gives $Q = 0$ and
\begin{equation}
P^{\prime\prime}_{n} +4t\left(\frac{b_1}{t^{2} -1} +
\frac{md_1}{mt^{2}-1}\right)P^{\prime}_{n} + G(t)P_n = 0,    \label{e27}
\end{equation}
where \\
\begin{eqnarray*}
G(t) = \frac{t^{2}(4b_{1}^{2} - 2b_{1} + 1/4)+1/2 -
  2b_{1}}{(t^{2}-1)^2} +\frac{(mt)^{2}(4d_{1}^{2} - 2d_{1} + 1/4)+m/2 -
  2md_{1}}{(mt^{2}-1)^2}  &&  \\ + \frac{2E -2j(j+1)mt^{2}
  +(16mb_{1}d_{1} - 1)mt^{2}}{2(t^{2}-1)(mt^{2}-1)} .\\
\end{eqnarray*}
 For each set of residues given in table I, this differential equation is
equivalent to a system of $n$ linear equations, for the
coefficients of different powers of $t$ in $P_{n}(t)$. The energy
eigenvalues are obtained by setting the corresponding determinant
equal to zero. We illustrate this process by obtaining the
eigenvalues and eigenfunctions explicitly for the case $j = 2$

\section{Lam{\'e} potential with $j=2$ }

   Using the procedure described in the previous section, we obtain
the eigenvalues and  the eigenfunctions for the supersymmetric
potential$^{\ref{khare}}$
\begin{equation}
V_{-}(x) = 6m \, \sn^{2}x - 2m -2 +2\delta,       \label{e28}
\end{equation}
where $\delta =  \sqrt{1-m+m^2}$. This potential is same as the Lam{\'e}
potential in \eqref{e2}{} with $j = 2$, except for an additive constant
$2\delta -2m -2$, which has been added to make the lowest energy equal
to zero.
The equation for $\chi(t)$ is,
\begin{equation}
\chi^{2} +\frac{\chi}{dt} + \frac{(mt)^{2}+2m}{4(1-mt^{2})^{2}}+\frac{t^2
  +2}{4(1-t^2)^{2}} +\frac{2E+4(m+1-\delta)-13mt^2}{2(1-t^2)(1-mt^2)} = 0.
  \label{e29}
\end{equation}
From \eqref{e23}{}, the number of moving poles $n$, for $j =2$ are
\begin{equation}
n = 3 - 2b_1 -2d_1,     \label{e30}
\end{equation}
where the values of $b_1$ and $d_1$ are obtained from \eqref{e18b}{}
and \eqref{e18c}{}. For each set of $b_1$, $d_1$ and
$n$, one can write the form and the number of linearly independent
solutions by substituting $j=2$, which gives $N=1$ in table I.
Hence for the case $j=2$, the form  and the number of solutions is as given in
table II.
%table II   
%\vskip0.1cm
%\hspace{1.0in} %\hspace{2in}
%\begin{table}[p]
%\caption{no of moving poles and number of solutions for $j=2$  }
%\begin{tabular}{|c|c|c|c|c|c|}
%\hline 
%\multicolumn{6}{|c|}
 %  \\  \hline
%\multicolumn{1}{|c|}{$set$} &\multicolumn{1}{|c|}{$b_1$}
%&\multicolumn{1}{|c|}{$d_1$} &\multicolumn{1}{|c|}{n}
%&\multicolumn{1}{|c|}{$\psi$ in terms of x}
%&\multicolumn{1}{|c|}{Number of solutions} \\
%\cline{6}
%\multicolumn{1}{c|}{$\psi$ in terms of x}
%&&&&&\multicolumn{1}{|c|}{$j=2N$}
%&\multicolumn{1}{|c|}{$j=2N$ }
% \hline
%$1$ & $1/4$ & $1/4$ & $2$ & $P_{2}(\sn x)$ &  $2$    \\
%$2$ & $3/4$ & $1/4$ & $j-1$ & $cnxP_{1}(\sn x)$  & $1$  \\
%$3$ & $1/4$ & $3/4$ & $j-1$ & $dnxP_{1}(\sn x)$  & $1$  \\
%$4$ & $3/4$ & $3/4$ & $j-2$ & $cnx dnxP_{0}(\sn x)$ & $1$  \\
%\hline
%\end{tabular}
%\end{table}
%\clearpage
     To get the unknown polynomial part $P_{n}(t)$ of the wave function
and the band edge energies, one needs to substitute the different
sets of $b_1$, $d_1$ and $n$ from table II, in the differential
equation
\begin{equation}
P^{\prime\prime}_{n} +4t\left(\frac{b_1}{t^{2} -1} +
\frac{md_1}{mt^{2}-1}\right)P^{\prime}_{n} + G(t)P_n = 0,
\label{e31}
\end{equation}
where
\begin{eqnarray*}
G(t) = \frac{t^{2}(4b_{1}^{2} - 2b_{1} + 1/4)+1/2 -
  2b_{1}}{(t^{2}-1)^2} +\frac{(mt)^{2}(4d_{1}^{2} - 2d_{1} + 1/4)+m/2 -
  2md_{1}}{(mt^{2}-1)^2} +  &&  \\
   \frac{2E +4(m+1-\delta)
  +(16b_{1}d_{1}-13)mt^{2}}{2(t^{2}-1)(mt^{2}-1)}.    % \label{e32}
\end{eqnarray*}

Thus for various sets of residues one has the band edge energies
and the wave functions as:

{\bf Set 1 : $b_1 = 1/4$ , $d_1 = 1/4 $, $ n = 2$} \\
Taking $P_2 =  At^{2}+Bt+C$, the parity constraint implies
\begin{equation}
B =0.    \label{e33}
\end{equation}
Substituting $b_1 , d_1$ and $ P_2$ in \eqref{e31}{}, one gets a
$2 \times 2$ matrix equation for $A$ and $C$ as follows

\begin{equation}
\left( \begin{array}{cc}
         E-2m-2-2\delta    &  -6m   \\
         2       &  E+2m+2-2\delta
\end{array}   \right)  \left( \begin{array}{c}
     A \\ C
\end{array}  \right)   = 0.    \label{e34}
\end{equation}
Equating the determinant of the matrix in \eqref{e34}{} to zero,
one gets the two values for energy as
 \begin{equation}
E_I = 4\delta ,\,\, E_{II} = 0,      \label{e35}
\end{equation}
which in turn give the polynomial as,
\begin{equation}
P_I = m + 1 - \delta - 3mt^2  ,\,\, P_{II} = m + 1 +\delta -3mt^2.   \label{e36}
\end{equation}
From \eqref{e26}{}, we see that, the band edge wave functions in
terms of $x$ will be
\begin{equation}
\psi_{I}(x) = m + 1 - \delta - 3m \, \sn^{2}x ,\,\, \psi_{II}(x) = m +
1 +\delta -3m \, \sn^{2}(x).  \label{e37}
\end{equation}

{\bf Set 2 : $b_1 = 3/4$ , $d_1 = 1/4 $, $ n = 1$} \\
  Taking $P_1 = t-t_k $ and substituting in \eqref{e31}{}, one gets
the band edge eigenvalue and $t_k$ as,
\begin{equation}
E = 2\delta - m +2  ,\,\,   t_k = 0.    \label{e38}
\end{equation}
The band edge wavefunction becomes
\begin{equation}
\psi(x) = \cn x \,\sn x.     \label{e40}
\end{equation}

{\bf Set 3 : $b_1 = 1/4$ , $d_1 = 3/4 $, $ n = 1$}\\
Proceeding in the same way as in set 2 one gets,
\begin{equation}
E = 2\delta +2m-1 ,\,\,  \psi(x) = \dn x \, \sn x.  \label{e41}
\end{equation}

{\bf Set 4 : $b_1 = 3/4$ , $d_1 = 3/4 $, $ n = 0$} \\
   In this case $P_0$ as a constant. One gets
the band edge energy and the wave function to be
\begin{equation}
E = 2\delta -m-1   ,\,\,   \psi(x) = \cn x \, \dn x.  \label{43}
\end{equation}
Thus one obtains five band edge wave functions and their
corresponding energies which agree with those given in
$^{\ref{khare}}$.

\section {Associated Lam{\'e} Potential}
The associated Lam{\'e} potential is ES when  $a = b =j$.
In this case the potential expression becomes
\begin{equation}
V(x) = j(j+1) m \left(\sn^{2}x +\frac{\cn^{2}x}{\dn^{2}x}\right).
\label{e43a}
\end{equation}
The QHJ equation with the associated Lam{\'e} potential is
\begin{equation}
q^2 + q^{\prime} + E -m j(j+1)\left(
\sn^{2}x+\frac{\cn^{2}x}{\dn^{2}x}\right) = 0.      \label{e45}
\end{equation}
Doing the change of variable $t = \sn x$ and proceeding in the same way
as in the case of general Lam{\'e} potential described in sec II,
one gets the equation for $\chi(t)$ as,
\begin{equation}
\chi^{2} +\frac{d\chi}{dt} + \frac{(mt)^{2}+2m(1-2j(j+1))}{4(1-mt^{2})^2}
  +\frac{2+t^2}{4(1-t^2)^2} +
  \frac{2E-mt^2(1+2j(j+1))}{2(1-t^2)(1-mt^2)}  = 0.   \label{e46}
\end{equation}

{\bf Singularity structure of $\chi(t)$ :} Similar to the Lam{\'e}
potential, $\chi(t)$ has fixed poles at $t = \pm 1$ and $t
= \pm 1/ \sqrt{m}$, along with $n$ moving poles in the entire complex $t$
plane and further we assume that, the point at infinity is an isolated
singularity. $\chi$ has no other singular points, except at $t = \pm
1$ and $t= \pm 1/ \sqrt{m}$. Following the same procedure, used in sec
$II$, one obtains the residues at $t =\pm 1$ as
\begin{equation}
b_1 = \frac{3}{4},\frac{1}{4} ,\,\, b_{1}^{\prime}
=\frac{3}{4},\frac{1}{4},    \label{e47}
\end{equation}
which is independent of $j$. The residues at $ t = \pm 1/
\sqrt{m}$ turn out to be
\begin{equation}
d_1 = \frac{3+2j}{4},\frac{1-2j}{4} ,\,\, d_1^{\prime} =
\frac{3+2j}{4},\frac{1-2j}{4},    \label{e48}
\end{equation}
 which are $j$ dependent.
Knowing the singularity structure of $\chi$, one can write it as
\begin{equation}
\chi =
\frac{b_{1}}{t-1}+\frac{b^{\prime}_{1}}{t+1}+\frac{d_{1}}{t-1/\sqrt{m}}+\frac{d^{\prime}_{1}}{t+1/\sqrt{m}}
+\frac{P^{\prime}_{n}}{P_n},    \label{e49}
\end{equation}
valid for all $t$ similar to the Lam{\'e} potential.

{\bf Behavior of $\chi$ at infinity :} The behavior of $\chi$ for large $t$, as fixed from the QHJ
equation \eqref{e46}{} is
\begin{equation}
\chi = \lambda_0 +\frac{\lambda_1}{t} +\frac{\lambda_2}{t^2} +....
\label{e50}
\end{equation}
and one gets the two values for $\lambda_1$ as
\begin{equation}
\lambda_1 = j+1 ,\,\, -j.    \label{e51}
\end{equation}
This result should agree with the large $t$ behavior of $\chi$
given by \eqref{e49}{}, which is
\begin{equation}
\chi = b_1 + b_{1}^{\prime}+  d_1 + d_{1}^{\prime} +n. \label{e52}
\end{equation}
As before, parity requirement implies $ b_1 =b_{1}^{\prime}$and  $
d_1 =d_{1}^{\prime}$ . Comparing the leading terms for large $t $,
in \eqref{e49}{} and \eqref{e52}{} we get,
\begin{equation}
2b_1 + 2d_1 +n = \lambda_1.      \label{e53}
\end{equation}
In this case both the values of $\lambda_1$ are allowed because $d_{1}$
and $d_{1}^{\prime}$ can take negative values unlike the case of Lam{\'e}
potential where $\lambda_1 = -j$ was ruled out. Thus one has two cases
$\lambda_1 = j +1$ and $\lambda_1 = -j$.

{\bf Case 1 : $\lambda = j+1$}

The values of $n$, which describe the number of zeros, for different
combinations of $b_1$ and $d_1$ values are given in table III.
%table III
%\vskip0.5cm \hspace{2in}
%\begin{tabular}{|c@{}|c@{}|c@{}|c@{}|p{2in|}}
%\begin{table}[p]
%\caption{values of $n$ for different combinations of $b_1$ and $d_1$}
%\begin{tabular}{|c|c|c|c|}
%\hline 
%%\multicolumn{4}{|c|}
%%\\ \hline
%\multicolumn{1}{|c|}{$set$}
%&\multicolumn{1}{c|}{$b_1$}
%&\multicolumn{1}{c|}{$d_1$}
%&\multicolumn{1}{c|}{$n$}
%\\ \hline
%    &       &                  &       \\
%$1$   &  $1/4$  & $(1-2j)/4$  & $2j$  \\
%$ 2$  &  $3/4$  & $(1-2j)/4$  & $2j-1$ \\
%&&&\\
%$3$   &  $1/4$  & $(3+2j)/4$  & -1  \\
%$4$   &  $3/4$  & $(3+2j)/4$  & -2  \\
%    &       &                  &      \\
% \hline
%\end{tabular}
%\end{table}
%\clearpage
%\vskip0.5cm
The sets 3 and 4 give $ n = -1 $ and $n=-2$ respectively and will not
be considered,  as $n$ should be greater than or equal to 0. The sets
1 and 2 will give positive 
values of $n$ only if $j$ is positive. Hence, this table is used
when $j$ is positive.

{\bf Case 2 : $ \lambda_{1} = -j$} \\
 The $n$ values for various  sets of $b_1$ and $d_1$ with $\lambda_{1}
 = -j$   are  listed in table IV .
%The requirement $n > 0$ implies that $j$ must be negative.
%table IV
%\vskip0.5cm
%\hspace{2in}
%\begin{table}[p]
%\caption{Various  sets of $b_1$ and $d_1$ values are listed below}
%\begin{tabular}{|c|c|c|c|}
%\hline 
%\multicolumn{4}{|c|}
%\\ \hline
%\multicolumn{1} {|c|} {$set$}
%&\multicolumn{1} {c|} {$b_1$}
%&\multicolumn{1} {c|} {$d_1$}
%&\multicolumn{1} {c|} {$n$}
%\\ \hline
%1 & $1/4$ & $ (1-2j)/4$ & $-1$ \\
%2 & $3/4$ & $ (1-2j)/4$ & $-2 $\\
%3 & $1/4$ & $ (3+2j)/4$ & $-2j-2$ \\
%&&&\\
%4 & $3/4$ & $ (3+2j)/4$ & $-2j-3$ \\
% \hline
%\end{tabular}
%\end{table}
%\clearpage
%\newpage
%\vskip0.5cm 
As in the previous case of $ \lambda_1 = j+1 $, we will not
consider the first two sets in the above table.\\ 

Note that the potential is invariant under the transformation $ j
\rightarrow -j-1$. From the two tables III and IV, we see that two
different values 
of $j$, one positive ($j = j^{\prime}
>0$) and another negative ($j = -j^{\prime} -1 < 0 $) leads to the
same expression for the potential described by \eqref{e3}{}. These
two values also lead to the same answers for the wave functions
and eigenvalues. Hence it is sufficient to restrict $j$ to
positive values alone. Thus using the two sets of combinations
given in table III and the expressions for the wave functions given
in \eqref{e26}{} one gets the following explicit forms and the
number of solutions given in  table V. 
%\vskip0.5cm
%Table V
%\hspace{2.0in}
%\begin{table}[p]
%\caption{Form of the wavefunctions for the Associated Lame{\'e} potential}
%\begin{tabular}{|c|c|c|c|}
%\hline 
%\multicolumn{4}{|c|}{Table V}
% \hline
%\multicolumn{1}{|c|}{$set$} &\multicolumn{1}{c|}{form of $\psi$}
%&\multicolumn{2}{|c|}{No of solutions}   \\  \cline{3-4}
%&&\multicolumn{1}{|c|}{$j=2N$} &\multicolumn{1}{|c|}{$j=2N+1$}
%\multicolumn{{2}{c@{\,\vline\,}{$j=2N$}  &  $j = 2N+1$ &
%\\ \hline
%    &                                      &      &         \\
%$1$ & $\frac{cnxP_{2j-1}(\sn x)}{(dnx)^{j}}$ & $2N$ & $ 2N+1$ \\
%&&&\\
%\\
%$2$ & $\frac{P_{2j}(\sn x)}{(dnx)^{j}}$ & $2N+1$ & $ 2N+2$ \\
%    &                                 &        &          \\
%\hline
%\end{tabular}
%\end{table}

%\vskip0.5cm

In the next section, we obtain explicit expressions for the wavefunctions and
energies for $j =1$.

\section{ Associated Lam{\'e} potential with $j = 1$}
We perform the calculation with the supersymmetric potential
\begin{equation}
  V_{-}(x) =  2m \,\sn^{2}x +2m \frac{\cn^{2}x}{\dn^{2}x} - 2- m + 2 \sqrt{1-m},
  \label{e54}
\end{equation}
corresponding to $j =1$. Proceeding in the same way  as
in the previous sections, the equation for $\chi(t)$ is found to be
\begin{equation}
\chi^{2} +\frac{d\chi}{dt}+\frac{(mt)^{2}-6m}{4(mt^{2}-1)^2}
+\frac{t^{2}+2}{4(t^{2}-1)^2} +
\frac{2E+4+2m-4\sqrt{1-m}-5mt^{2}}{2(1-mt^{2})(1-t^{2})} = 0
\label{e55}
\end{equation}
and the form for $\chi(t)$ is
\begin{equation}
\chi =
\frac{b_{1}}{t-1}+\frac{b^{\prime}_{1}}{t+1}+\frac{d_{1}}{t-1/\sqrt{m}}+\frac{d^{\prime}_{1}}{t+1/\sqrt{m}}
+\frac{P^{\prime}_{n}}{P_n}.    \label{e56}
\end{equation}
Substituting \eqref{e56} in \eqref{e55}{}, one is left with the
differential equation
\begin{equation}
P^{\prime\prime}_{n} +4t\left(\frac{b_1}{t^{2} -1} +
\frac{md_1}{mt^{2}-1}\right)P^{\prime}_{n} + G(t)P_n = 0,
\end{equation}   \label{e57}
where
\begin{eqnarray*}
G(t) = \frac{t^{2}(4b_{1}^{2} - 2b_{1} + 1/4)+1/4 -
  2b_{1}}{(t^{2}-1)^2} +\frac{(mt)^{2}(4d_{1}^{2} - 2d_{1} + 1/4)+-3m/2 -
  2md_{1}}{(mt^{2}-1)^2}+  &&  \\
\frac{2E+4+2m-4\sqrt{1-m}+(16b_{1}d_{1}-5)mt^{2}}{2(t^{2}-1)(mt^{2}-1)}.
 %   \label{e58}
\end{eqnarray*}

Substituting $j =1$ in table III,  one gets the band edge
eigenfunctions and eigenvalues from set 1 and set 2  as follows.\\
{\bf Set 1 : $b_1 = 3/4$ , $d_1 = -1/4 $, $ n = 1$} \\
 This combination gives only one solution with
\begin{equation}
E = 2-m+2\sqrt{1-m}  ,\,\,   \psi(x) = \frac{\cn x \, \sn x}{\dn x}.
\label{e59}
\end{equation}

{\bf Set 2 : $b_1 = 1/4$ , $d_1 = -1/4 $, $ n = 2$} \\
This combination gives two solutions. The band edge energies are
\begin{equation}
E_I = 0  ,\,\, E_{II} = 4\sqrt{1-m}     \label{e61}
\end{equation}
and the corresponding wave functions are
\begin{equation}
\psi_{I}(x) = \frac{1}{m}\left(\dn x +\frac{\sqrt{1-m}}{\dn x}\right)  ,\,\,
  \psi_{II}(x) = \frac{1}{m}\left(\dn x -\frac{\sqrt{1-m}}{\dn x}\right),   \label{e62}
\end{equation}
which match with the wave functions in$^{\ref{khare}}$.

\section{conclusions}

    In this paper, we have studied the ES periodic potentials of the elliptic
class  namely the Lam{\'e} and the associated Lam{\'e} potentials. Using the
QHJ formalism we have obtained here the general form of the band edge
wave functions for these potentials, with the potential parameters
taking only integer values . Also, the band edge eigenfunctions and
the energy eigenvalues have been obtained explicitly for $j =2$, for the
Lam{\'e} potential and $p = q = 2$, for the associated Lam{\'e} potential.  In
the process, we have studied $p$, the logarithmic derivative of the wave
function  for these elliptic potentials and have found an interesting
singularity structure exhibited by $p$ in the complex domain. 

    The QHJ formalism has been successful in obtaining the energy eigenvalues
and the wave functions for a large variety of ES models and periodic
potentials. The most important steps in obtaining the solutions using QHJ
formalism have been, the choice of the change of variable and the
ability to guess the right singularity structure of the QMF. It should
be noted here that, our approach has been to assume a singularity
structure of the QMF as simple as possible.  Once a 
correct choice has been found, one could obtain the solutions in a
most straight forward fashion even if the resulting equations were
not of a simple form. 
   From our earlier study of ES$^ {\ref{sree}}$, QES$^{\ref{geo}}$ and
the present study of periodic potentials, some interesting features of
these models appear to be correct. All the models whether ES, QES or
periodic potentials, which we have studied so
far, share a common property that `the QMF', $\chi$, becomes a rational function after a
suitable change of variable. Therefore, in the finite complex plane it has
finite number of moving poles, described by the parameter $n$. Our
assumption that the point at infinity is an isolated singular point is
equivalent to the condition that the QMF has finite number of moving poles.

    There are interesting differences in details of the singularity 
structure for different potential models. For the ES models,
each value of $n$ corresponds to an energy level and an eigenfunction with
only $n$ real zeros. These  zeros correspond to the $n$ moving poles of the
QMF and are confined to the  classical region only.  For the QES
models, $n$ appears  as a parameter in the potential and
of the infinite possible states only $n$ states can be obtained
analytically. The QMF for all these states will have the same number of
moving poles which are both real and complex. Correspondingly 
the  wavefunctions for each state will have the same number  $(\equiv
n)$ of zeroes, of which the number of real zeros are in accordance with
the oscillation  theorem. The number of real zeros increases as one
goes from the lowest state to the highest state possible.

 On the other hand, the exactly solvable {\em periodic} potential
 models show quite a 
different kind of distribution of the moving poles in the complex plane. Like
QES models, the moving poles of the QMF consist of both real and complex
poles. But unlike the QES models, where the number of poles of QMF for  all 
the known states is same, for ES Periodic potential models  one finds 
groups of solutions, with number of moving poles remaining same within a 
group but varying from one group to another group of solutions. This
point becomes clear from the form of the solutions listed in table I
for the Lam{\'e} potential. Different groups of solutions are
precisely the different sets of the solutions listed in table I. In
each set, the degree of the polynomial $P_{n}(\sn x)$  is same and
the total number of zeros of the band edge wave functions in the
 interval $0 \leq x < 2K(m)$ are
same. For example, for the $N+1$ linear independent solutions belonging
 to set 1 the 
total number of zeros of the wavefunctions will be $j$. However for 
the solutions belonging to set 2 there will be $j-1$ zeros of
$P_{n}(\sn x)$ and  one zero corresponding to $ \cn x$, thus a total of
$j$ zeros. For all the sets the total number of zeros are given in the
last column of table I. However the number of real zeros in a group
increases with  increasing energy and correspondingly the number of
complex zeros will  decrease. For all values of $j$, we will have
utmost four such groups. A similar pattern is observed for ES
associated  Lam{\'e} potential.

   When one computes the residues at the fixed pole one gets two
solutions due to the quadratic nature of the QHJ equation. For
the bound state problems, both ES and QES considered in the earlier
papers,$^{\ref{sree},\ref{geo}}$ the square integrability of the wave function
had to be insisted. This allowed us to pick the right residue, out of the
two values, which gave the physically acceptable wave function, but in the
present case of periodic potential models, there is no such
restriction and, both values of the 
residues lead to acceptable solutions. The only restriction on the
possible combination of the residues comes from the requirements of
parity.
          
   To conclude we note that the new QHJ equation, obtained after a 
change of variable, involves rational functions in the 
independent variable. In order to make progress, we made crucial 
assumption that there are finite number of moving poles implying that 
the point at infinity is an isolated singular point. This important assumption 
has been used for all the models studied and its consistency with other
equations has given useful restrictions on the acceptable solutions. 
Our present study completes this series of investigation of ES models
from the QHJ approach. We now have a good starting point to take up
models which are not ES or QES. For these models, making a
proper guess about the singularity structure of the QMF does not appear
to be possible.  A study
of the distribution of the moving poles in the complex plane may have
a relation to the classical properties of the system. Such a study is
interesting in its own right 
and can be best done numerically. We hope that here also, the
established results in the complex variable theory will provide a
useful scheme to obtain numerical, possibly  approximate analytical
solutions for the wave functions and energies.

%\pagebreak

%{\bf Acknowledgments :}

{\bf References}

\begin{enumerate}

\item {\label{opl}} C. Fort, F. S. Cataliotti, L. Fallani, F.
Ferlaino, P. Maddaloni and M. Inguscio, Phys. Rev. Lett {\bf 90}, 140405 (2003).

\item {\label{kit}} C. Kittel, {\it Introduction to Solid state
  Physics}, 191 (Fifth edition, Wiley Eastern Limited, New Delhi).

\item {\label{kuz}} H. Li and D. Kuznetsov, Phys. Rev. Lett. {\bf 83}, 1283
(1999).

\item {\label{d1}} G. V. Dunne and M. Shifmann, Annals of Phys {\bf 299},
  143 (2002). 

\item {\label{pav}} P. Ivanov, J. Phys A : Math. Gen. {\bf 34}, 8145
  (2001); quant - ph/0008008. 

\item {\label{osc}} O. Rosas - 0rtiz, Published in  Preccedings of the IV workshop on {\it Gravitational and Mathematical Physics}, N. Bret\'{o}n {\it et. al} (Eds),Chapala Jalisco, Mexico (2001), math - ph/ 0302189.

\item {\label{khare}} A.Khare and U. Sukhatme, J. Math. Phys. {\bf
  40}, 5473 (1999); quant - ph/9906044.

\item {\label{d2}} G. V. Dunne and J. Fienberg, Phys. Rev. D. {\bf 57} 1271 (1998).

\item {\label{bec1}} J. C. Bronski, L. D. Carr, B. Deconinck and
J. N. Kutz, Phys. Rev. Lett {\bf 86}, 1402 (2001).

\item {\label{sc}} F. L. Scarf, Phys. Rev. {\bf 112}, 1137 (1958) .

\item {\label{asis}} A. Ganguly, J. Math. Phys. {\bf 43}, 1980 (2002); math - ph/0207028.

\item {\label{tka}} V. M. Tkachuk and O. Voznyak, Phys. Lett. A
  {\bf301}, 177 (2002). 

\item {\label{ars}} F. M. Arscot, {\it Periodic Differential Equations}
   (Pergamon,Oxford, 1964).

\item {\label{mag}} W. Magnus and S. Wrinkler, {\it Hills
Equation} (Interscince Publishers, New York, 1966).

\item {\label{whit}} E. Whittaker and G. N. Watson, {\it A Course of Modern
  Analysis} (Cambridge Univ. Press, Cambridge, 1980).

\item {\label{fin}} F. Finkel, A. Gonzalez - Lopez, A. L. Maroto and
  M. A. Rodriguez, Phys. Rev. D {\bf 62} (2002) 103515; hep - ph/0006117.

\item {\label{kan}} R. Kantowski and R. C. Thomas; astro - ph/0011176.

\item {\label{mok}} O. I. Mokhov, math. DG/0201224.

\item {\label{ mar}} M. Bouhmadi - Lopez, L. J. Garay and P. F. Gonzalez
  - Diaz, Phys. Rev. D {\bf 66}, 083504 (2002); gr - qc/0204072.

\item {\label{ale}} A. V. Razumov and M. V. Saveliev, Published in
  Proceedings of the International Conference {\it Selected Topics of
  Theorietical and Modern Mathematical Physics (SIMI - 96)}, Tbilisi,
  Georgia (1996); XXX AR Xiv : solv - int/9 612004.

\item {\label{dav}} D. J. Fernandez C, B. Mielnik, O. Rosas - Ortiz and
 B. F. Samsonov, J. Phys. A {\bf 35}, 4279 (2002); quant - ph/0303051.

\item {\label{bac}} I. Bacus, A. Brandhuber and K. Sfetsos,  Contribution
  to the TMR meeting {\it Quantum Aspects of Guage Theories, Supersymmetry and Unification}, Paris (1999);  hep - th/0002092.

\item {\label{brac1}} M. Brack, M. Mehta and K. Tanaka, J. Phys. A
  {\bf34}, 8199 (2001); nlin. CD/0105048.

\item {\label{gli}} M. D. Glinchuk, E. A. Eliseev and V. A. Stephanovich;
  cond - matt/0103083. 

\item {\label{brac2}} M. Brack, S. N. Fedotkin, A. G. Magner and
  M. Mehta, J. Phys A {\bf 36}, 1095 (2003); nlin. CD/0207043. 

\item {\label{uday}}  A. Khare and U. sukhatme, J. Math. Phys {\bf 42}
  5652 (2001); quant - ph/0105044.

 \item {\label{tur1}} A. V. Turbiner, Commun. Math. Phys. {\bf118},467 (1988).

\item {\label{shif}} M. A. Shifmann, Contemp. Math. {\bf 160}, 237 (1998).

\item {\label{tur2}} A. V. Turbiner, J. Phys. A. {\bf 22} LI (1989).

\item {\label{ Gonz}} F. Finkel, A. Gonlzalez - Lopez and  M. A. Rodriguez, J. Math. Phys. {\bf 37}, 3954 (1996).

\item {\label{alh}} Y. Alhasad , F. G{\"u}rsey and F. Iachello,
  Phys. Rev. Lett. {\bf 50}, 873 (1983).

\item {\label{lea}} R. A. Leacock and M. J. Padgett,
  Phys. Rev. Lett. {\bf 50} 3 (1983).

\item {\label{pad}} R. A. Leacock and M. J. Padgett, Phys. Rev. D  
 {\bf 28}, 2491 (1983).

\item {\label{bh1}}R. S. Bhalla, A.  K. Kapoor and P. K. Panigrahi,
 Am. J. Phys. {\bf 65}, 1187 (1997).

\item {\label{bh2}} R. S. Bhalla,  A. K. Kapoor and P. K
 Panigrahi, Mod. Phys. Lett. A. {\bf 12}, 295 (1997).

\item {\label{bh3}} R. S. Bhalla, A. K. Kapoor and P. K.
 Panigrahi, Phys. Rev. A. {\bf 54},951  (1996).

\item {\label{bh4}} R. S. Bhalla, A. K. Kapoor and P. K.
 Panigrahi, Int. J. Mod. Phys. A. {\bf 12}, 1875 (1997).

\item {\label{geo}} K. G. Geogo, S. Sree Ranjani and A. K. Kapoor, J. Phys
  A : Math. Gen. {\bf 36}, 4591 (2003); quant - ph/0207036.

\item {\label{sree}} S. Sree Ranjani, K. G. Geojo, A. K Kapoor and
  P. K. Panigrahi, quant - ph/0211168.

\item {\label{ince}} E. L. Ince, {\it Ordinary Differential equations}
  (Dover publications Inc, New York, 1956).

\item {\label{chu}} R. V. Churchill and J. W. Brown, Complex Variables
  and Applications ( McGraw -  Hill Publishing Company, New York).

\item {\label{han}} H. Hancock, Theory of Elliptic Functions, Dover
  Publications, Inc. New York, (1958).

\end{enumerate}

%\listoftables

\begin{table}[p]
\caption{Form of the wave functions $\psi(x)$, number of linear independent solutions and the total
  number of zeros of $\psi(x)$ for the general Lam{\'e} potential $V(x)= j(j+1)m \, \sn^{2}x$. }
\begin{tabular}{|c|c|c|c|c|c|c|c|}
\hline
%\multicolumn{7}{|c|}
%   \\  \hline
\multicolumn{1}{|c|}{set}
&\multicolumn{1}{|c|}{$b_1$}
&\multicolumn{1}{|c|}{$d_1$}
&\multicolumn{1}{|c|}{n}
&\multicolumn{1}{|c|}{$\psi$ in terms of x}
%&\multicolumn{1}{|c|}{Total zeros of $\psi(x)$}
&\multicolumn{2}{|c|}{Number of solutions}& Total zeros  \\
\cline{6-7}
%\multicolumn{1}{c|}{$\psi$ in terms of x}
&&&&&\multicolumn{1}{|c|}{$j=2N+1$}
&\multicolumn{1}{|c|}{$j=2N$ }
&\multicolumn{1}{|c|}{of $\psi(x)$} 
 \\ \hline
$1$ & $1/4$ & $1/4$ & $j$ & $P_{j}(\sn x)$ & $N+1$ & $N+1$ & $j$    \\
$2$ & $3/4$ & $1/4$ & $j-1$ & $\cn x P_{j-1}(\sn x)$ & $N+1$ & $N$ & $j$  \\
$3$ & $1/4$ & $3/4$ & $j-1$ & $\dn x P_{j-1}(\sn x)$ & $N+1$ & $N$ & $j-1$  \\
$4$ & $3/4$ & $3/4$ & $j-2$ & $\cn x \dn xP_{j-2}(\sn x)$ & $N$ & $N$
 & $j-1$  \\
\hline
\end{tabular}
\end{table}
%\clearpage

%\newpage
%\clearpage
%table II   
\newpage
%\vskip0.1cm
%\hspace{1.0in} %\hspace{2in}
\begin{table}[p]
\caption{ The form of the wave functions and the number of
linear independent  solutions for the Lam{\'e} potential with $j=2$,
$V_{-}   = 6m \sn ^{2}x - 2m -2 +2 \delta $  where $\delta = \sqrt{1-m
+m^2}$.}
\begin{tabular}{|c|c|c|c|c|c|}
\hline 
%\multicolumn{6}{|c|}
 %  \\  \hline
\multicolumn{1}{|c|}{$set$} &\multicolumn{1}{|c|}{$b_1$}
&\multicolumn{1}{|c|}{$d_1$} &\multicolumn{1}{|c|}{n}
&\multicolumn{1}{|c|}{$\psi$ in terms of x}
&\multicolumn{1}{|c|}{Number of LI solutions} \\
%\cline{6}
%\multicolumn{1}{c|}{$\psi$ in terms of x}
%&&&&&\multicolumn{1}{|c|}{$j=2N$}
%&\multicolumn{1}{|c|}{$j=2N$ }
 \hline
$1$ & $1/4$ & $1/4$ & $2$ & $P_{2}(\sn x)$ &  $2$    \\
$2$ & $3/4$ & $1/4$ & $1$ & $\cn x P_{1}(\sn x)$  & $1$  \\
$3$ & $1/4$ & $3/4$ & $1$ & $\dn x P_{1}(\sn x)$  & $1$  \\
$4$ & $3/4$ & $3/4$ & $0$ & $\cn x \dn x P_{0}(\sn x)$ & $1$  \\
\hline
\end{tabular}
\end{table}
%\clearpage

%\clearpage
%table III
\newpage
%\vskip0.5cm \hspace{2in}
%\begin{tabular}{|c@{}|c@{}|c@{}|c@{}|p{2in|}}
\begin{table}[p]
\caption{Values of $n$ for different combinations of $b_1$ and $d_1$
for $\lambda_1 = j+1$ for the associated Lam{\'e} potential $V(x) =
j(j+1)m( \sn ^{2}x + \cn ^{2}x / \dn ^{2}x)$.}
\begin{tabular}{|c|c|c|c|}
\hline 
%\multicolumn{4}{|c|}
%\\ \hline
\multicolumn{1}{|c|}{$set$}
&\multicolumn{1}{c|}{$b_1$}
&\multicolumn{1}{c|}{$d_1$}
&\multicolumn{1}{c|}{$n$}
\\ \hline
%    &       &                  &       \\
$1$   &  $1/4$  & $(1-2j)/4$  & $2j$  \\
$ 2$  &  $3/4$  & $(1-2j)/4$  & $2j-1$ \\
%&&&\\
$3$   &  $1/4$  & $(3+2j)/4$  & -1  \\
$4$   &  $3/4$  & $(3+2j)/4$  & -2  \\
%    &       &                  &      \\
 \hline
\end{tabular}
\end{table}
\vskip0.5cm

%\clearpage
%table IV
\newpage
%\vskip0.5cm
%table4
%\hspace{2in}
\begin{table}[p]
\caption{Various  sets of $b_1$ and $d_1$ for $\lambda_1 = -j$ for
associated Lam{\'e} potential $V(x) =
j(j+1)m( \sn ^{2}x + \cn ^{2}x / \dn ^{2}x)$.}  
\begin{tabular}{|c|c|c|c|}
\hline 
%\multicolumn{4}{|c|}
%\\ \hline
\multicolumn{1} {|c|} {$set$}
&\multicolumn{1} {c|} {$b_1$}
&\multicolumn{1} {c|} {$d_1$}
&\multicolumn{1} {c|} {$n$}
\\ \hline

1 & $1/4$ & $ (1-2j)/4$ & $-1$ \\
2 & $3/4$ & $ (1-2j)/4$ & $-2 $\\
3 & $1/4$ & $ (3+2j)/4$ & $-2j-2$ \\
%&&&\\
4 & $3/4$ & $ (3+2j)/4$ & $-2j-3$ \\

 \hline
\end{tabular}
\end{table}

%\clearpage
%table 5
\newpage
\begin{table}[p]
\caption{Form of the wavefunctions for the associated Lam{\'e}
potential $V(x) =
j(j+1)m( \sn ^{2}x + \cn ^{2}x / \dn ^{2}x)$ for positive values of $j$.}
\begin{tabular}{|c|c|c|c|}
%\hline 
%\multicolumn{4}{|c|}{Table V}
 \hline
\multicolumn{1}{|c|}{$set$} &\multicolumn{1}{c|}{form of $\psi$}
&\multicolumn{2}{|c|}{No of solutions}   \\  \cline{3-4}
&&\multicolumn{1}{|c|}{$j=2N$} &\multicolumn{1}{|c|}{$j=2N+1$}
%\multicolumn{{2}{c@{\,\vline\,}{$j=2N$}  &  $j = 2N+1$ &
\\ \hline
    &                                      &      &         \\
$1$ & $\frac{\cn x P_{2j-1}(\sn x)}{(\dn x)^{j}}$ & $2N$ & $ 2N+1$ \\
&&&\\
%\\
$2$ & $\frac{P_{2j}(\sn x)}{(\dn x)^{j}}$ & $2N+1$ & $ 2N+2$ \\
    &                                 &        &          \\
\hline
\end{tabular}
\end{table}

%\clearpage

\end{document}